\documentclass[showpacs,preprintnumbers,english,twocolumn,amsmath,amssymb]{revtex4-1}
\usepackage[T1]{fontenc}
\usepackage[latin1]{inputenc}
\usepackage{graphicx}
\usepackage{amssymb}
\usepackage{verbatim}
\usepackage{epic,eepic}
\usepackage{babel}

\begin{document}

\title{Non-Abelian behavior of $\alpha$ Bosons in cold symmetric nuclear matter }

%\footnote{Funded in part by DOE and NSF-REU Program}}
\author{ Hua Zheng$^{a,b)}$ and Aldo Bonasera$^{a,c)}$}
%\email{ruslan.magana@nucleares.unam.mx}
\affiliation{
a)Cyclotron Institute, Texas A\&M University, College Station, TX 77843, USA;\\
b)Physics Department, Texas A\&M University, College Station, TX 77843, USA;\\
c)Laboratori Nazionali del Sud, INFN, via Santa Sofia, 62, 95123 Catania, Italy.}

%\author{Hua Zheng$^a)$}
%\email{hzheng@comp.tamu.edu}
%\affiliation{Cyclotron Institute, Texas A\&M University, College Station, TX 77843, USA}

%\author{Aldo Bonasera$^{a,c)}$}
%\email{abonasera@comp.tamu.edu}
%\affiliation{Cyclotron Institute, Texas A\&M University, College Station, TX 77843, USA}
%\affiliation{Laboratori Nazionali del Sud, INFN, via Santa Sofia, 62, 95123 Catania, Italy.}

%\date{September 8, 2010}

\begin{abstract}
The ground state energy of infinite symmetric nuclear matter is usually described by strongly interacting nucleons obeying the Pauli exclusion principle.  We can imagine a  unitary transformation which 
groups four non identical nucleons (i.e. with different spin and isospin) close in coordinate space.  Those nucleons, being non identical, do not obey the Pauli principle, thus their relative momenta are negligibly small
(just to fulfill the Heisenberg principle).  Such a cluster can be identified with an $\alpha$ boson.  But in dense nuclear matter, those $\alpha$ particles still obey the Pauli principle since are constituted
of Fermions.  The ground state energy of nuclear matter  $\alpha$ clusters is the same as for nucleons, thus it is degenerate.  We could think of $\alpha$ particles as vortices which can now braid, for instance making $^8Be$ which leave the ground state energy
unchanged.  Further braiding to heavier clusters ($^{12}C$, $^{16}O$..) could give a different representation of the ground state at no energy cost.  In contrast d-like clusters (i.e. N=Z odd-odd nuclei, where N and Z are the neutron and proton number respectively)
 cannot describe the ground state of nuclear matter and can be formed at high excitation energies (or temperatures) only. We show that even-even , N=Z, clusters could be classified as non-Abelian states of matter.  As a consequence an $\alpha$ condensate in nuclear matter 
might be hindered by the Fermi motion, while it could be possible a condensate of $^8Be$ or heavier clusters.

\end{abstract}

\pacs{ 25.70.Pq}
%42.50.Lc,64.70.Lg
\maketitle

%\section{Introduction}
Nuclei are composed of strongly interacting nucleons.  In some conditions they can also be described as $\alpha$ clusters \cite{ring}.  Thus Fermions in one case and Bosons in the other. However, this distinction into different symmetry particles is not 
the entire story.  In fact for Bosons like nuclei an important constraint is the Pauli principle which makes their wave functions antisymmetric in the exchange of two particles, thus they behave as Fermions.  Recently,  interest has grown for states of non definite symmetry
which are classified as non-Abelian\cite{stern}. Those are  composite particles which do not follow the Boson or the Fermion statistics.  Properties of those composites are\cite{stern,8}:\\
1) The ground state of the system is degenerate, not based on symmetry arguments of the wave function,  and there exist an energy gap to the first excited state.\\
2) Exchanging two (quasi)particles  does not result simply in a change of sign such as for Fermions or Bosons. More importantly those exchanges might lead or not the system from a possible ground state to another depending on the order of the exchanges. This
ordering dependence thus is non commutative or non-Abelian.  \\
To see if those properties could be fulfilled by nuclear systems, let us consider the ground state of $^{12}C$.  This can be well described by strongly interacting nucleons, but we can also think of grouping four $\it {non-identical}$ nucleons
to make an $\alpha$ particle, thus the ground state is described as 3$\alpha$ clusters.  However, to get the correct ground state energy is not possible if we group the nucleons into deuteron like systems, or if we group the particles into four $\it {identical}$ Fermions. 
The latter cases will result in higher
energies.  Of course a combination of $\alpha$ particles and nucleons might be possible as well to recover the ground state energy, even though such combinations might be more probable for heavier systems.  The first excited state 
of $^{12}C$ is at about 7 MeV, the breaking into  3 alphas (Hoyle state)\cite {ring}.  Thus the first two properties of non-Abelian systems might be recovered in nuclear systems but there are other conditions to be fulfilled which we
will discuss later when going into the details of an infinite symmetric nuclear matter system. Other important informations can be obtained from heavy ion collisions at beam energies around the Fermi energy.  Fragmentation of those nuclei, especially in collisions of $\alpha$-
cluster candidates, e.g. $^{40}Ca+^{40}Ca$  etc., show a large production of $\alpha$ particles compared to nucleons, but ${\it not}$ of d-clusters\cite{1,10}.\\
 Infinite nuclear matter is an idealized case of a system made of neutrons and protons where, to avoid singularities, the Coulomb force is
conveniently 'switched  off', i.e. the properties of nuclei are corrected for the Coulomb energy and the limit for the mass number A going to infinity is taken.  Also we will be considering symmetric nuclear matter, i.e. the number of neutrons N is equal to 
the number of protons Z: N=Z.  This simplifies the discussion as far as the symmetry energy is concerned and more importantly the possibility of clustering of matter into tritons, helions, etc..Those composite are Fermions and their combination could as well 
represent the ground state of symmetric nuclear matter but, as for deuterium, the resulting energy is higher than the well established energy of symmetric nuclear matter $E/A=-15 MeV$.  Thus we will neglect those states in the present work and concentrate
on Boson-like composites only.
We could write the energy per particle $N_x$ of nuclear matter as \cite{rus}:
\begin{equation}
 E/N_x=S\tilde \varepsilon_f\tilde\rho^\frac{2}{3}+\sum_{n=1}^k\frac{A_n}{n+1}\tilde \rho^n- BE/N_x
\label{2.7}
\end{equation}
where the first term refers to the average kinetic energy of a free Fermi gas with $\tilde \varepsilon_f=3/5 \varepsilon_f= 22.5 MeV$. S=1 or 0 for Fermions or Bosons respectively, but there will be some exceptions as discussed below. 
$N_x=A$, the number of nucleons, when considering the system made of A nucleons.  Otherwise $N_x=A/A_{cl}$  where $A_{cl}$ is the mass number of a cluster (2 for d, 4 for $\alpha$ etc.) when it is assumed that nuclear matter is made of  clusters.
 The other terms are due to  potential interactions and correlations. The $n=1$ term  is obtained by taking into account  the interaction between pair of particles, and the subsequent terms must involve the interactions between groups of three, four, etc., particles. 
The coefficients $A_n$ in the expansion, eq. (\ref{2.7}) are called first, second,  third, etc., virial coefficients\cite{8}. The last term $BE/N_x$ is the binding energy of the cluster corrected by the Coulomb energy which is simply given by $V_c=\frac{3}{5} 1.44 \frac{Z_{cl}^2}{A_{cl}^{1/3}}$.
Such a term (which is zero for nucleons) simply reflects the fact that when the density goes to zero, the energy per particle becomes the binding energy per particle of the cluster. $\tilde \rho=\frac{\rho}{\rho_0}$ is the reduced density of the system, $\rho_0=\frac{A/A_{cl}}{V_0}$ is the ground state density, $V_0$ its volume.
The equation of state (EOS) for nucleons has been derived in \cite{rus} and we will adopt the $CCS\delta 3$ parametrization here. It includes the possibility of a quark-gluon plasma at high densities which is not important for this work since we will be discussing mainly results at
densities lower than the ground state density.  We stress that any 'nucleons' EOS gives similar results if they fulfill the conditions discussed below.  This EOS will be our reference to compare to the cases where clustering is assumed.\\
In order to fit the parameters entering eq.(1) we will impose the conditions\cite{1,10,5,shlomo}:
\begin{equation}
\begin{array}{lll}
a)&E/N_x=-15 A_{cl}\text{MeV}\\
b)&P=0&\text{at}\hspace{0.5cm}\rho=\rho_0\\
c)&K=225 A_{cl}\text{MeV}
\end{array}
\label{2.6}
\end{equation}
where the pressure P is given by:
\begin{equation}
 P=\rho^2\frac{\partial (E/A)}{\partial \rho}
\label{2.2}
\end{equation}
and the (isothermal) compressibility,  is defined in nuclear physics  as:
\begin{equation}
 K=9\frac{\partial P}{\partial\rho}\bigg|_{\rho=\rho_0,T=0}
\label{2.3}
\end{equation}
Those three conditions imply that we have enough informations to include third order terms in the expansion in eq.(1), i.e. four body forces.  From physical arguments we need $A_3\ge 0$ otherwise the system will collapse to infinite density\cite{rus}.  The clusters binding energy is obtained from experimental data 
corrected by Coulomb as discussed above.
\begin{figure}
\centering
\includegraphics[width=1.15\columnwidth]{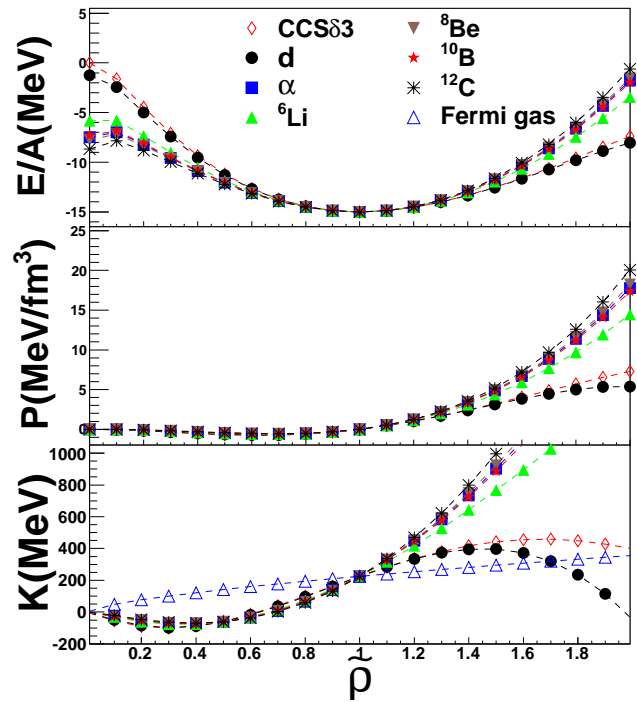} 
\caption[]{(Color online) From top to bottom, energy per particle, pressure and compressibility vs. reduced density for cluster matter (full symbols) or the $CCS\delta 3$ EOS (open symbols).  Different cluster types are indicated in the figure. The compressibility of a non interacting 
Fermi gas is also given for reference (open triangles).}
\label{Fig1}
\end{figure}
The Fermi energy has to be included for d, $\alpha$ since they are built of non identical particles.  When  two or more of those clusters are within a volume in phase space comparable to $h^3$, the Planck's constant, their constituents must fulfill the
Pauli principle.  On the other hand heavier clusters have their own Fermi motion thus the first term in eq.(1) should be zero for those particles.  However,
we might consider those heavier systems made up of d or $\alpha$ clusters as well and include the Fermi motion in eq.(1).  We will discuss the case when heavier cluster are made of $\ nucleons$ with their own Fermi motion below. Now we consider heavier clusters as a 
'braid' of $\alpha$ particles (or d particles for odd- odd clusters).  Using the conditions above, eq.(2), we can solve the three equations (1,3,4) and obtain the relevant parameters.  The result is plotted in figure 1 where we see that, by construction, all different clusters
give the correct properties of infinite nuclear matter, eq.(2). For completeness we report below the values of the parameters for some cases:
\begin{equation}
\begin{array}{lll}
a)A_1=-360;A_2=180; A_3=0\text{MeV for $\alpha$ particles}.\\
b)A_1=-255;A_2=202.5; A_3=-50\text{MeV for $d$ particles}.\\
\end{array}
\label{2.61}
\end{equation}

The first important difference is that odd-odd nuclei give $A_3 < 0$, e.g. eq.(5b), thus their EOS are unphysical. It is also seen in the figure that at high densities, even-even nuclei have a larger energy than odd-odd and more importantly that matter
would collapse into deuterons if their EOS would be correct.  Notice the value of $A_3=0$ in eq.(5a)  which implies that four body forces are not important for $\alpha$ clusters.  This might be a mere coincidence or another indication
that nuclear matter can indeed be treated as $\alpha$ clusters.  We will discuss the meaning of $A_3=0$ for other cases below.
At lower densities the compressibility becomes negative  for all systems, which means that the system is unstable.  In particular, at zero temperature, the system would prefer clustering into particles of higher binding energy.  
The situation would be different at finite temperature where the entropy would favor nucleons rather than clusters.  A discussion for finite temperature is also interesting and will be the subject of a  following work. Thus, as we see from figure (1), the ground state of nuclear matter could well described by different
$spin=0$ clusters and it is degenerate.  Those Bosons still obey the Pauli principle, thus their properties also at finite temperatures, are different from when they are at zero density.   Of course when we go from nuclear densities to atomic densities, the Fermi motion of those
particles is negligible and the Bosons properties are recovered, i.e we can have a superfluid helium liquid.  The ground state is thus degenerate and stable against small perturbations. As we have seen
we have to compress or decompress the system in order that different clusterizations are selected: at high densities clusters will dissolve into nucleons (and at even higher densities into quarks and gluons), at lower densities, nucleons will coalesce into clusters of higher and higher binding energy.
Thus in those conditions the system might prefer one configuration respect to the others. Those features might give rise to first order phase transitions at low densities or cross-over at higher densities.  \\
Clusterization  of particles with $spin=1$ or higher are not possible as we discussed above but we could ask the question of what should be the energy of nuclear matter if they were build of such clusters.  To have an estimate of such an energy, we could leave the energy per particle in
eq.(1) as a free parameter and impose that $A_3=0$ and $A_2>0$.  Solving the simple system of equations will give the results plotted in figure 2.
 \begin{figure}
\centering
\includegraphics[width=1.15\columnwidth]{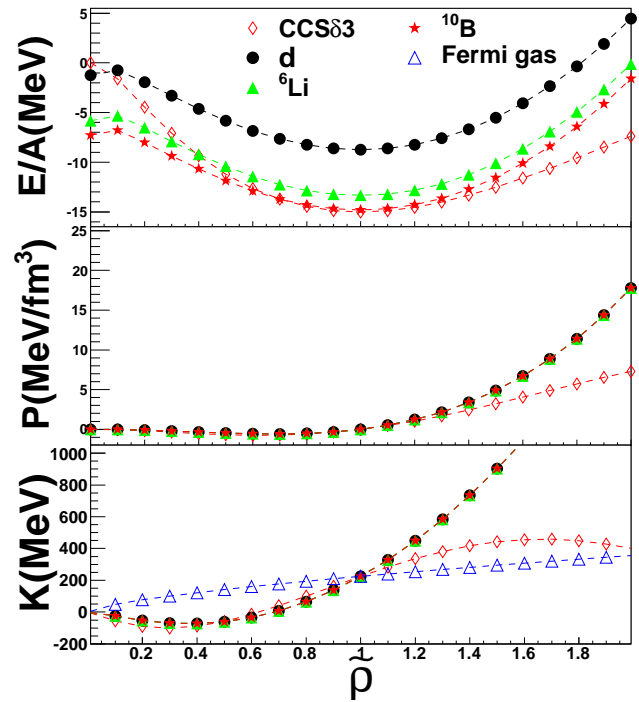}
\caption[]{(Color online) Same as in figure 1 for odd-odd clusters. Symbols as in figure 1. }
\label{Fig2}
\end{figure} 
As we see the energy of such systems would be higher than the ground state energy of infinite nuclear matter but will become comparable for heavier odd-odd clusters.  They could appear at lower densities since their energy is larger than those of nucleons, but they will have to compete 
with even-even clusters. At higher densities they are not favored by energy considerations and will break into nucleons.  At  even higher densities, nucleons will break into quarks and gluons.\\
Now we can consider the situation where nucleons could group into clusters of 8, 10, 12...particles. In this case those clusters contain identical particles, thus the constituents have a Fermi motion.  The cluster is now a Boson like and forcing two of such Bosons
into the microscopic phase space volume does not call for antisymmetrization if we neglect the difference of the Fermi momenta of those clusters and the Fermi momentum of infinite nuclear matter.  Of course such an approximation becomes more and more exact with increasing
size of the cluster.  As a result we can now neglect the Fermi motion in eq.(1) and calculate the EOS for heavier clusters.  The results are plotted in figure 3 and compared to the $CCS\delta3$ EOS.  
   \begin{figure}
\centering
\includegraphics[width=1.15\columnwidth]{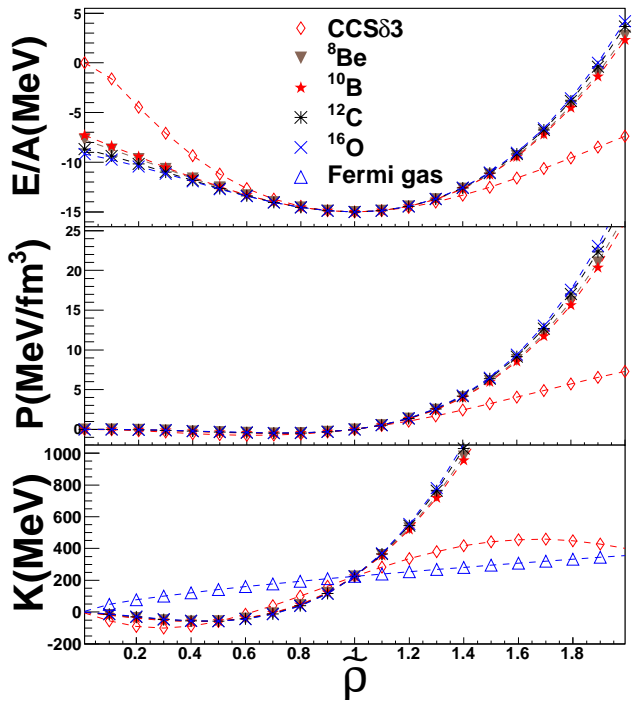} 
\caption[]{(Color online) Same as figure 1 for bosons.}
\label{Fig3}
\end{figure}  
Now the coefficient $A_3>0$ for all plotted particles including $^{10}B$ which was not the case when such a cluster was considered as a d-like system.  Also in this case the basic properties of the infinite nuclear matter can be recovered which means
that the ground state is degenerate even in this representation.  Very importantly,  a Bose condensate is now possible starting with the $^8Be$ cluster . We stress that we would obtain a state of larger energy if we would group 8,10,12.. $identical$ nucleons.  Thus the transformations which lead to such states are not commutative.
It is interesting to note that in all the physical situations addressed in this work, the compressibility is very similar, figures 1-3, and systematically lower than the Fermi gas compressibility.  We expect the difference between the Fermi gas result and the strongly interacting cases to decrease with increasing temperatures.
Thus the Fermi gas compressibility could be a good approximation when deriving temperatures and densities from multiplicity and quadrupole fluctuations in dynamical nuclear systems \cite{trho}.\\
The ideas discussed in this paper could be investigated experimentally in heavy ion collisions.  The possibility of describing nuclei as $\alpha$ clusters is quite well established\cite{ring}.  Some progress has been done recently \cite{15a} in  cluster formation
in low density nuclear matter at finite temperatures. Efforts are being made to search for boson condensates in nuclei\cite{15a,horiuchi}.  A confirmation of some of the ideas discussed here would be the determination of a condensate of $^8Be$ or higher cluster sizes as discussed above.  
Different choices of the interaction, for instance strong momentum dependent potentials \cite{15a}, might give an $\alpha$ condensate as well differently from our predictions.  To distinguish among different models a possibility would be to compare multiplicity fluctuations
of protons and $\alpha$ \cite{trho}.  Those fluctuations  are quite different for Fermions or Bosons \cite{8} thus could distinguish between the two cases.  If the fluctuations turn out to be the same ( apart some Coulomb effects that could be determined by comparing neutrons,
tritons and helions multiplicity fluctuations) it could be a signature that $\alpha$ particles do not form a condensate in nuclei as suggested by some authors\cite{horiuchi}.  Thus an experimental investigation of multiplicity fluctuations in the same experimental conditions could be quite interesting, for instance in $^{40}Ca+^{40}Ca$
 at beam energies around the Fermi energy or below. Similar ideas could be applied to pions and kaons and other bosons produced in ultra relativistic heavy ion collisions.  Quarks and gluons could play the role of the nucleons in nuclear matter while hadrons would be clusters of mixed symmetries.

%\begin{acknowledgments} We thank prof. J.Natowitz for 'animated' discussions.
%\end{acknowledgments}

\end{document}